\documentclass[twocolumn,aps,pra,superscriptaddress,tightenlines]{revtex4-2}
\usepackage{amsmath}
\usepackage{amssymb}
\usepackage{graphicx}
\usepackage{epsfig}
\usepackage{epstopdf}
\usepackage{color}
\usepackage{epsfig}
\usepackage{amsfonts}
\usepackage{xcolor}
\usepackage{mathtools}
\usepackage[colorlinks,citecolor=blue]{hyperref}
\usepackage{multirow}
\usepackage{footnote}

\def\narrowtext{\par\global\columnwidth20.5pc
	\global\hsize\columnwidth\global\linewidth\columnwidth
	\global\displaywidth\columnwidth}
\makeatletter
\renewcommand{\maketag@@@}[1]{\hbox{\m@th\normalsize\normalfont#1}}%
\makeatother
\begin{document}
	
	\title{Nonreciprocal routing induced by chirality in an atom-dimer waveguide-QED system}
	
	\author{Shi-Yu Liu}
	\email{These authors contributed equally to this work.}
	\affiliation{Key Laboratory of Low-Dimensional Quantum Structures and Quantum Control of Ministry of Education, Key Laboratory for Matter Microstructure and Function of Hunan Province, Department of Physics and Synergetic Innovation Center for Quantum Effects and Applications, Hunan Normal University, Changsha 410081, China} 
	
	\author{Lin-Lin Jiang}
	\email{These authors contributed equally to this work.}
	\affiliation{Key Laboratory of Low-Dimensional Quantum Structures and Quantum Control of Ministry of Education, Key Laboratory for Matter Microstructure and Function of Hunan Province, Department of Physics and Synergetic Innovation Center for Quantum Effects and Applications, Hunan Normal University, Changsha 410081, China} 
	
	\author{Hai Zhu}
	\affiliation{Key Laboratory of Low-Dimensional Quantum Structures and Quantum Control of Ministry of Education, Key Laboratory for Matter Microstructure and Function of Hunan Province, Department of Physics and Synergetic Innovation Center for Quantum Effects and Applications, Hunan Normal University, Changsha 410081, China}
	
	\author{Jie-Qiao Liao}
	\affiliation{Key Laboratory of Low-Dimensional Quantum Structures and Quantum Control of Ministry of Education, Key Laboratory for Matter Microstructure and Function of Hunan Province, Department of Physics and Synergetic Innovation Center for Quantum Effects and Applications, Hunan Normal University, Changsha 410081, China}
	\affiliation{Institute of Interdisciplinary Studies, Hunan Normal University, Changsha 410081, China}
	\affiliation{Hunan Research Center of the Basic Discipline for Quantum Effects and Quantum Technologies, Hunan Normal University, Changsha 410081, China}
	
	\author{Jin-Feng Huang}
	\email{Contact author: jfhuang@hunnu.edu.cn}
	\affiliation{Key Laboratory of Low-Dimensional Quantum Structures and Quantum Control of
		Ministry of Education, Key Laboratory for Matter Microstructure and Function of Hunan Province, Department of Physics and Synergetic Innovation Center for Quantum Effects and Applications, Hunan Normal University, Changsha 410081, China}
	\affiliation{Institute of Interdisciplinary Studies, Hunan Normal University, Changsha 410081, China}
	\affiliation{Hunan Research Center of the Basic Discipline for Quantum Effects and Quantum Technologies, Hunan Normal University, Changsha 410081, China}

	\begin{abstract}
The implementation of quantum routers is an important and desired task in quantum information science, since quantum routers are important components of quantum networks. Here we propose a scheme for implementing single-photon routers in a waveguide-QED system, which consists of two coupled two-level atoms coupled to two waveguides to form a four-port quantum device. We obtain the exact analytical expressions of the single-photon scattering amplitudes using the real-space method. By taking into account or disregarding the propagating time of photons between two coupling points, we consider the system working in the Markovian and non-Markovian regimes, respectively. In addition, we introduce the chiral coupling, which breaks the symmetry of the waveguide model, to manipulate the transmission of single photons. We find that when the system works in the non-Markovian regime, the single photon can be transmitted on demand by adjusting the asymmetry coefficient. More interestingly, the complete single-photon routing in this device does not require an ideal chiral coupling,
loosening the photon transport conditions. This work should motivate studies concerning the nonreciprocal and chiral quantum devices in the waveguide-QED platform.
	\end{abstract}
	\maketitle
	\narrowtext
	
	\section{Introduction\label{introduction}}
	
The quantum router, as an important quantum device, has wide potential applications in quantum information processing and quantum networks~\cite{Kimble2008}.In particular, the development of efficient and controllable quantum routers is critical for
realizing large-scale quantum networks~\cite{chou2007functional}.~Usually, quantum
routing refers to the controllable transmission of quantum states from a starting node to a target node through designated paths in a quantum network. Due to this fascinating function, much recent attention has been paid to both the theoretical~\cite{Aoki2009,Chudzicki2010,zhou2013quantum,yan2014single, zhu2019single,yang2018phase,shomroni2014all,Huang201301,Huang201302,yin2022,Zhu2025,Gheeraert2020,Solano2023} and experimental~\cite{hoi2011demonstration,Reiserer2014,Wang2021,Webber2020,Yuan2015,cao2017implementation,Guimond2020,Kannan2022}research on quantum routers. In realistic quantum information processing, the information is often encoded in single-photon pulses, making the routing of single photons a fundamental requirement. Notably,
recent studies have adopted monochromatic wave analysis for quantum routing~\cite{zhou2013quantum,lu2014single,zhu2019single,Li2015,tan2019,yang2018}. So far, quantum routers have been realized in various physical platforms, such as optical
systems~\cite{Wang2020}, cavity-QED~\cite{zhou2013quantum,lu2014single,zheng2011,yan2016,yan2017}, trapped ions~\cite{Monroe2013}, superconducting circuits~\cite{Kemp2011,Liu2006,Niskanen2007,Xiong2015,Liao2009}, coupled-resonator waveguide~\cite{Cheng2012,Shi2011,Zhou2018,Werra2013,Zheng2012, Liao201001,Liao201002}, and quantum dots~\cite{Bentham2015}.
	
	Nonreciprocal transmission of photons~\cite{Mitsch2014,Petersen2014,Söllner2015,Young2015,Lodahl2015,Zheng2024,Xu2017},  describing
	photons propagating asymmetrically through a medium or
	interface, provides a physical mechanism for the implementation of quantum routers. In nonreciprocal propagation, the
	transmission and reflection probabilities of photons depend on the propagation direction. In classical systems, nonreciprocity
	is typically achieved using elements like isolators~\cite{Jalas2013,Fan2009} and circulators~\cite{Sliwa2015}, which usually rely on magnetic fields or nonlinearity. In quantum systems, achieving nonreciprocity~\cite{Leonardo2015} at the single-photon level can be realized through various approaches, such as utilizing nonlinear optical interactions~\cite{Metelmann2015,Shang2019}, chiral quantum optics~\cite{Lodahl2017,Grankin2018,guimond2016chiral,yan2016photonic,zhou2023chiral, yan2018tunable,Suárez-Forero2025}, synthetic magnetic fields~\cite{Fang2012}, and optomechanical systems~\cite{li2018tunable,Manipatruni2009,Hafezi2012,Xu2015}.
	In addition, the chiral waveguide-QED systems~\cite{Mitsch2014,Petersen2014,Söllner2015,Young2015}, breaking the time-reversal symmetry, can provide an efficient platform for realizing nonreciprocal propagation of photons. Various models describing the couplings between different atoms with the chiral waveguides have been proposed, such as the $\Delta$- and $\Lambda$-type three-level atoms~\cite{sundaresan2019interacting,song2020controlling,sanchez2016full,zhao2020single,Yan2018targeted} and giant atoms~\cite{kockum2018decoherence,du2022giant,carollo2020mechanism,du2021single,kannan2020waveguide,Roccati2024}. To
	achieve perfect nonreciprocal routing, an ideal chiral-coupling condition is often required. However, the realization of ideal chiral couplings requires special waveguide design and precise quantum emitter position regulation, as well as loss, scattering, and coupling inhomogeneities in actual physical
	systems.
	
	In this work we study nonreciprocal routing of photons in a waveguide-QED system in which two coupled two-level atoms (TLAs) are chirally coupled to two one-dimensional
	infinite waveguides. This dual-waveguide dual-atom configuration constitutes a four-port system, enabling multipath routing and expanding beyond the limitations of singlewaveguide setups. Compared with single-atom routers, the dual-atom scheme introduces dipole-dipole coupling as a different controllable element. Moreover, our motivation is to investigate the fundamental physical mechanisms of single-photon routing by analyzing the energy spectrum, symmetry, interactions, and coupling structure. To clarify these inherent parameter dependences, we focus on the stationary scattering in this system.
	
Concretely, we adopt the real-space method to obtain
the exact analytical expressions of the single-photon scattering amplitudes in both the Markovian and non-Markovian
regimes. The results show that the scattering amplitudes of the
single photon depend on the phase shift between the coupling
points, the coupling strength between the two TLAs, and the
coupling strength between the TLAs and the waveguides. In
the Markovian regime, the phase shift is independent of the incident energy. We also find that the single-photon scattering is
nonreciprocal in the chiral-coupling case. In order to achieve
a perfect single-photon routing, we further investigate the
dependence of the scattering behavior on the chiral condition
and the dipole coupling strength between the two TLAs. In
the non-Markovian regime, we find that the scattering spectra
exhibit richer structures, with multiple peaks and staggered
dips, which are caused by quantum interference. In addition,
we find that the non-Markovian effect can enhance the nonreciprocity of single-photon scattering in the chiral-coupling
case. More interestingly, the present system does not need to
satisfy the ideal chiral-coupling condition to realize a nonreciprocal routing.
	
	The rest of this paper is organized as follows. In Sec. \ref{model_hamiltonian} we present the physical model and the Hamiltonians. In Sec. \ref{calculate amplitude} we calculate the scattering solution of single photon incident from either port 1 or port 2. In Sec. \ref{Markovian regime} we study the single-photon scattering under the chiral coupling in the Markovian regime, and realize the target routing of photons. In Sec.\ref{non-Markovian regime} we study the single-photon scattering in the nonMarkovian regime and discuss the realization of the target
	routing.  In Sec. \ref{DISCUSSIONS} we discuss the physical implementation
	of our scheme. We summarize in Sec. \ref{Conclusion}.
	
	\begin{figure}[tbp]
		\center
		\includegraphics[clip, width=8.5cm]{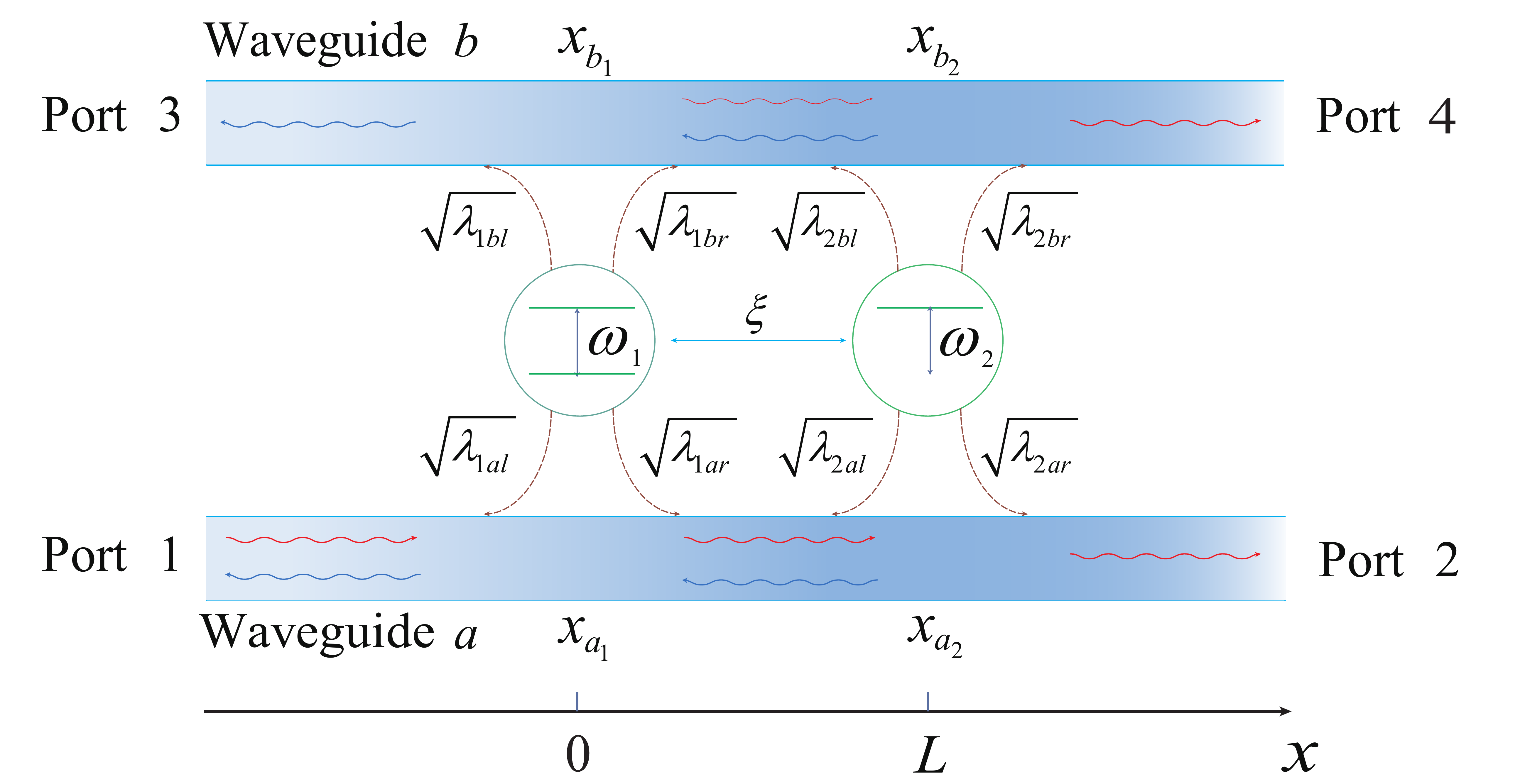}
		\caption{Schematic of the four-port waveguide-QED system. Two independent waveguides $a$ and $b$ interact
			with two TLAs, which are coupled with each other via the dipole interaction with the coupling strength $\xi$. The $i$th $(i=1,2)$ TLA couples to waveguides $j$ at the coupling point $x_{j_{i}}$. Here we take $x_{j_{1}}=0$ and $x_{j_{2}}=L$~$(j = a, b)$.}
		\label{Fig1_model}
	\end{figure}
	%\vspace{-8pt}
	\section{Physical Model and Hamiltonian}~\label{model_hamiltonian}
	We consider two coupled TLAs coupled commonly to two one-dimensional infinite waveguides $a$ and $b$, as shown in Fig.~\ref{Fig1_model}. The first TLA couples to the two waveguides $a$ and $b$ at positions $x_{a_{1}}=0$ and $x_{b_{1}}=0$, while
	the second TLA couples to both waveguides at $x_{a_{2}}=L$ and $x_{b_{2}}=L$. The two TLAs couple to each other via a dipole-dipole interaction with the coupling strength $\xi$.  The
	total Hamiltonian of the system reads $(\hbar=1)$
	\begin{eqnarray}
		\hat{H}=\hat{H}_{A}+\hat{H}_{F}+\hat{H}_{I},
	\end{eqnarray}%
	where
	\begin{eqnarray}
		\hat{H}_{A}\! &\!\!=\!\!&  \omega_{1}\left\vert e\rangle\!_{1 1}\!\langle e\right\vert +\omega_{2}\left\vert e\rangle\!_{2 2}\!\langle e\right\vert +\xi\left(\sigma_{1}^{+}\sigma_{2}^{-}+\sigma_{1}^{-}\sigma_{2}^{+}\right),   \nonumber \\
		\hat{H}_{F}\!&\!\!=\!\!&\sum_{j=a,b}i\upsilon_{g}\int\!dx\!\left[\hat{a}_{jl}^{\dag}(x)\frac{\partial}{\partial x}\hat{a}_{jl}(x)\!-\!\hat{a}_{jr}^{\dag}(x)\frac{\partial}{\partial x}\hat{a}_{jr}(x)\right],    \nonumber \\
		\hat{H}_{I}\!&\!\!=\!\!&\sum_{j=a,b}\!\int\! dx\delta(x)\!\left[\!\sqrt{\lambda_{1jl}}\hat{a}_{jl}(x)\!+\!\sqrt{\lambda_{1jr}}\hat{a}_{jr}(x)\right]\sigma_{1}^{+}   \nonumber \\
		&& \!+\!\sum_{j=a,b}\!\int\! dx\delta(x\!-\!L)\!\left[\!\sqrt{\lambda_{2jl}}\hat{a}_{jl}(x)\!+\!\sqrt{\lambda_{2jr}}\hat{a}_{jr}(x)\right]\sigma_{2}^{+}   \nonumber \\ 
		&& + \textrm{H.c.}. 
	\end{eqnarray}
	Here, $\hat{H}_{A}$ is the Hamiltonian of the two coupled TLAs. The $\omega_{i}$ is the transition frequency between the excited state $\left\vert e\right\rangle _{i}$ $(i=1,2)$ and the ground state $\left\vert g\right\rangle _{i}$ of the $i$th TLA, with $\sigma_{i}^{+}=(\sigma_{i}^{-})^{\dagger}=\left\vert e\rangle\!_{i i}\!\langle g\right\vert$ $(i=1,2)$ describing the transition of the TLA $i$. The $\xi$ term describes the dipole interaction between the two TLAs. The $\hat{H}_{F}$ is the free Hamiltonian of the two waveguides,
	where $\ensuremath{\hat{a}_{jr}^{\dagger}\left(x\right)=\left[\hat{a}_{jr}\left(x\right)\right]^{\dagger}}$
	and
	$\ensuremath{\hat{a}_{jl}^{\dagger}\left(x\right)=\left[\hat{a}_{jl}\left(x\right)\right]^{\dagger}}$
	are the creation operators for creating a right- and left-moving photon in
	waveguide $j$ $\left(j=a,b\right)$ at position $x$ with group velocity
	$\upsilon_{g}$, respectively. The Hamiltonian $\hat{H}_{I}$ describes the interactions between the TLAs and the waveguides.  In addition, $\sqrt{\lambda_{ijl}}$ and $\sqrt{\lambda_{ijr}}$  $\left(i=1,2;j=a,b\right)$ are the coupling strengths between the TLA $i$ and the waveguide $j$ for left- or right-moving photon, respectively.
	
	To understand the physical process of the single-photon routing, we diagonalize $\hat{H}_{A}$ by solving the eigen-equation
	$\hat{H}_{A}\left\vert E_{n} \right\rangle =E_{n}\left\vert E_{n}\right\rangle$ for $n=1,2,3,4$.
	Then we obtain the following four eigenstates 
	\begin{eqnarray}\label{lambda}
		\left\vert E _{1}\right\rangle &=& \left\vert e\right\rangle _{1}\left\vert e\right\rangle _{2},  \notag \\
		\left\vert E _{2}\right\rangle&=& \cos\left(\psi/2\right)\left\vert e\right\rangle _{1}\left\vert g\right\rangle _{2}+\sin\left(\psi/2\right)\left\vert e\right\rangle _{2}\left\vert g\right\rangle _{1}, \notag \\
		\left\vert E_{3}\right\rangle &=& -\sin\left(\psi/2\right)\left\vert e\right\rangle _{1}\left\vert g\right\rangle _{2}+\cos\left(\psi/2\right)\left\vert e\right\rangle _{2}\left\vert g\right\rangle _{1}, \notag \\
		\left\vert E_{4}\right\rangle  &=& \left\vert g\right\rangle _{1}\left\vert g\right\rangle _{2},
	\end{eqnarray}
	with the corresponding eigenenergies
	\begin{eqnarray}
		E_{1}&=&\omega_{1}+\omega_{2}, \notag \\ E_{2}&=&\frac{1}{2}[\omega_{1}+\omega_{2}+\sqrt{4\xi^{2}+(\omega_{1}-\omega_{2})^{2}}], \notag \\ 
		E_{3}&=&\frac{1}{2}[\omega_{1}+\omega_{2}-\sqrt{4\xi^{2}+(\omega_{1}-\omega_{2})^{2}}], \notag \\
		E_{4}&=&0. 
	\end{eqnarray}
   Here, we introduce the mix angle $\psi$ defined by $\tan\psi=2\xi/(\omega_1-\omega_2)$.
	When $\omega_{1}=\omega_{2}$, it is found that $E_{1}-E_{2}=E_{3}-E_{4}$ and $E_{1}-E_{3}=E_{2}-E_{4}$.
	
	Since the total excitation number of the system is a conserved quantity, we can restrict the system within the single-excitation subspace to study the single-photon scattering process. In this subspace, the eigenstate $\left\vert \Phi\right\rangle $ of the system can be written as
	\begin{eqnarray} \label{Eq5}
		\left\vert \Phi\right\rangle  & \!\!=\!\!\! & \ensuremath{\ensuremath{\int dx\left[\phi_{ar}(x)a_{ar}^{\dag}(x)+\phi_{al}(x)a_{al}^{\dag}(x)\right]\left\vert \varnothing\right\rangle \left\vert g\right\rangle _{1}\left\vert g\right\rangle }_{2}} \notag \\
		&  & +\int dx\left[\phi_{br}(x)a_{br}^{\dag}(x)+\phi_{bl}(x)a_{bl}^{\dag}(x)\right]\left\vert \varnothing\right\rangle \left\vert g\right\rangle _{1}\left\vert g\right\rangle _{2} \notag \\
		&  &+u_{1}\left\vert \varnothing\right\rangle \left\vert e\right\rangle _{1}\left\vert g\right\rangle _{2}+u_{2}\left\vert \varnothing\right\rangle \left\vert e\right\rangle _{2}\left\vert g\right\rangle _{1}.
	\end{eqnarray}%
	Here, $\phi_{jr}(x)$ and $\phi_{jl}(x)$ are the wave functions of the single right- and left-propagating photon in the waveguide $j$. The coefficient $u_{i}$ $(i=1,2)$ is the probability amplitude for the $i$th TLA to be in its excited state. The state $\left\vert \varnothing\right\rangle$ denotes the vacuum state of the waveguides. We note that the form of $\left\vert \Phi\right\rangle$ is applicable for the cases whenever the single photon is input from either the right or the left end of waveguide $a$ or $b$.
	
	We first consider the photon input from port 1, i.e., the left end of waveguide $a$, as shown in Fig.~\ref{Fig1_model}. Then the wave functions $\phi_{jr}\left(x\right)$ and $\phi_{jl}\left(x\right)$ can be expressed as
	\begin{align} \label{Eq6}
		\phi_{ar}(x)\!&=\!  e^{ik_{a}\!x}\left[\Theta(-x)+\!t_{12a}\Theta(x)\Theta(L-x)\!+t_{a}\Theta(x-L)\right],  \nonumber\\
		\phi_{al}(x)\!&=\!  e^{-ik_{a}\!x}\left[r_{a}\Theta(-x)+r_{12a}\Theta(x)\Theta(L-x)\right], \nonumber\\
		\phi_{br}(x)\!&=\!  e^{ik_{b}\!x}\left[t_{12b}\Theta(x)\Theta(L-x)+t_{b}\Theta(x-L)\right], \nonumber \\
		\phi_{bl}(x)\!&=\! e^{-ik_{b}\!x}\left[r_{b}\Theta(-x)+r_{12b}\Theta(x)\Theta(L-x)\right].
	\end{align}
	Here, $\Theta(x)$
	is the Heaviside step function, and $k_{j}$ is the wave vector in waveguide $j$.~The term $e^{ik_{j}x}t_{12j}\Theta(x)\Theta(L-x)$ in $\phi_{jr}$ is the wave function of the single right-propagating photon between $x_{j_{1}}\!=0$ and $x_{j_{2}}\!=L$ in waveguide $j$ with the transmission amplitude $t_{12j}$, while the term $e^{-ik_{j}x}r_{12j}\Theta(x)\Theta(L-x)$ in $\phi_{jl}$ is the wave function of the single left-propagating photon between $x_{j_{1}}=0$ and $x_{j_{2}}=L$ with the reflection amplitude $r_{12j}$. Moreover, $t_{j}$ ($r_{j}$) is the transmission (reflection) amplitude for $x>L$ ($x<0$) in waveguide $j$. As the photon propagates in waveguide $j$, it will carry a phase term $e^{ik_{j}x}$ between the two coupling points of waveguide $j$.

	To study the nonreciprocity of the single-photon routing, we consider the photon initially input from port 2, i.e., the right end of waveguide $a$, as shown in Fig.~\ref{Fig1_model}. In this case, compared with the photon input from port 1, the phase term between the two coupled points of waveguide $j$ becomes $e^{-ik_{j}x}$.
	The corresponding wave functions $\phi_{jr}\left(x\right)$ and $\phi_{jl}\left(x\right)$ for the photon incident from port 2 become
	\begin{eqnarray}
		\widetilde{\phi}_{ar}(x)\!\!\!&=&\!\!\!e^{ik_{a}x}\left[r_{12a}\Theta(x)\Theta(L\!-\!x)+r_{a}\Theta(x\!-\!L)\right], \notag\\
		\widetilde{\phi}_{al}(x)\!\!\!&=&\!\!\! e^{-ik_{a}x}\left[t_{a}\Theta(-x)\!+\!t_{12a}\Theta(x)\Theta(L\!-\!x)\!+\!\Theta(x\!-\!L)\right], \notag\\
		\widetilde{\phi}_{br}(x)\!\!\!&=&\!\!\! e^{ik_{b}x}\left[r_{12b}\Theta(x)\Theta(L\!-\!x)+r_{b}\Theta(x\!-\!L)\right],  \notag\\
		\widetilde{\phi}_{bl}(x)\!\!\!&=&\!\!\! e^{-ik_{b}x}\left[t_{b}\Theta(-x)+t_{12b}\Theta(x)\Theta(L\!-\!x)\right].
	\end{eqnarray}
	Here, to distinguish the wave functions input from port 1, we add a tilde in $\phi_{jr}$ and $\phi_{jl}$. Similarly, the term $e^{ik_{j}x}r_{12j}\Theta(x)\Theta(L-x)$ in $\widetilde{\phi}_{jr}$ is the wave function of the single right-propagating photon between $x_{j_{1}}=0$ and $x_{j_{2}}=L$ in waveguide $j$ with the reflection amplitude $r_{12j}$, while the term $e^{-ik_{j}x}t_{12j}\Theta(x)\Theta(L-x)$ in $\widetilde{\phi}_{jl}$ is the wave function of the single left-propagating photon between $x_{j_{1}}=0$ and $x_{j_{2}}=L$ in waveguide $j$ with the transmission amplitude $t_{12j}$.
	Accordingly, $t_{j}$ ($r_{j}$) is the transmission (reflection) amplitude for $x<0$ ($x>L$) in waveguide $j$.
	
	\section{Single-photon transmission and reflection amplitudes}~\label{calculate amplitude}
	To quantitatively characterize the single-photon routing, we need to calculate the transmission and reflection amplitudes. We first consider the case of a single photon incident from port 1 with energy $\varepsilon $. By solving the eigenequation $\hat{H}\left\vert \Phi\right\rangle =\varepsilon\left\vert \Phi\right\rangle $, we can get the equations
	\begin{eqnarray}\label{Eq8}
		\varepsilon\phi_{al}(x)\!\!\!&=&\!\!\! \delta(x)\!\sqrt{\lambda_{1al}}u_{1}\!+\!\delta(x\!-\!L)\!\sqrt{\lambda_{2al}}u_{2}\!+\!i\upsilon_{g}\frac{\partial}{\partial x}\phi_{al}(x),\notag \\
		\varepsilon\phi_{ar}(x)\!\!\!&=&\!\!\! \delta(x)\!\sqrt{\lambda_{1ar}}u_{1}\!+\!\delta(x\!-\!L)\!\sqrt{\lambda_{2ar}}u_{2}\!-\!i\upsilon_{g}\frac{\partial}{\partial x}\phi_{ar}(x),\notag \\
		\varepsilon\phi_{bl}(x)\!\!\!&=&\!\!\! \delta(x)\!\sqrt{\lambda_{1bl}}u_{1}\!+\!\delta(x\!-\!L)\!\sqrt{\lambda_{2bl}}u_{2}\!+\!i\upsilon_{g}\frac{\partial}{\partial x}\phi_{bl}(x),\notag \\
		\varepsilon\phi_{br}(x)\!\!\!&=&\!\!\! \delta(x)\!\sqrt{\lambda_{1br}}u_{1}\!+\!\delta\left(x\!-\!L\right)\!\sqrt{\lambda_{2br}}u_{2}\!-\!i\upsilon_{g}\frac{\partial}{\partial x}\phi_{br}\!\left(x\right).\notag \\
	\end{eqnarray}
	By integrating Eq.~(\ref{Eq8}) around $x=0$ and $x=L$, respectively, we obtain
	\begin{eqnarray} \label{Eq9}
		i\upsilon_{g}\left(r_{12a}-r_{a}\right)+u_{1}\sqrt{\lambda_{1al}} & = & 0, \notag \\
		-i\upsilon_{g}\left(t_{12a}-1\right)+u_{1}\sqrt{\lambda_{1ar}}& = & 0, \notag \\
		i\upsilon_{g}\left(r_{12b}-r_{b}\right)+u_{1}\sqrt{\lambda_{1bl}}& = & 0, \notag \\
		-i\upsilon_{g}t_{12b}+u_{1}\sqrt{\lambda_{1br}}& = & 0, \notag \\
		i\upsilon_{g}\left(-e^{-ik_{a}L}r_{12a}\right)+u_{2}\sqrt{\lambda_{2al}}& = & 0, \notag \\
		-i\upsilon_{g}\left(e^{ik_{a}L}t_{a}-e^{ik_{a}L}t_{12a}\right)+u_{2}\sqrt{\lambda_{2ar}}& = & 0, \notag \\
		i\upsilon_{g}\left(-e^{-ik_{b}L}r_{12b}\right)+u_{2}\sqrt{\lambda_{2bl}}& = & 0, \notag \\
		-i\upsilon_{g}\left(e^{ik_{b}L}t_{b}-e^{ik_{b}L}t_{12b}\right)+u_{2}\sqrt{\lambda_{2br}}& = & 0,      
	\end{eqnarray}
	and
	\begin{eqnarray} \label{Eq10}
		\Delta_{1}u_{1} & \!\!=\!\! & \frac{1}{2}\left[\sqrt{\lambda_{1ar}}(1+t_{12a})\!+\!\sqrt{\lambda_{1al}}(r_{a}+r_{12a})\right]+\xi u_{2}  \notag \\
		&&+ \frac{1}{2}\left[\sqrt{\lambda_{1br}}t_{12b}+\sqrt{\lambda_{1bl}}(r_{b}+r_{12b})\right], \notag \\  
		\Delta_{2}u_{2}& \!\!=\!\! &\frac{1}{2}\left[\sqrt{\lambda_{2ar}}(t_{12a}\!+\!t_{a})e^{ik_{a}L}\!\!+\!\!\sqrt{\lambda_{2al}}r_{12a}e^{-ik_{a}L}\right]\!+\!\xi u_{1} \notag \\
		&& \!+\!\frac{1}{2}\left[\sqrt{\lambda_{2bl}}r_{12b}e^{-ik_{b}L}\!+\!\sqrt{\lambda_{2br}}(t_{12b}\!+\!t_{b})e^{ik_{b}L}\right].  \notag \\
	\end{eqnarray}
	Here, we introduce the detuning $\Delta_{i}=\varepsilon-\omega_{i}~(i\!=1, 2)$ between the frequency of the photon and that of the $i$th TLA. For simplicity, we assume $\omega_{1}=\omega_{2}=\omega_{e}$; then $\Delta_{1}=\Delta_{2}=\Delta$. We also assume that the coupling strengths between the left-moving (right-moving) photon in both waveguides and the two TLAs are identical, namely, $\sqrt{\lambda_{ijl}}=\sqrt{\lambda_{1}}$ and $\sqrt{\lambda_{ijr}}=\sqrt{\lambda_{2}}~(i=1,2$ and $j=a,b)$. By solving Eqs.~(\ref{Eq9}) and~(\ref{Eq10}), we can obtain the transmission and reflection amplitudes corresponding to the single photon incident from port 1 as 
	\begin{subequations} \label{0_ta}
		\begin{align}
			t_{a}&=-\dfrac{2e^{2i\varphi}B+2ie^{i\varphi}\sqrt{c_{1}}\xi-2\sin\varphi\sqrt{c_{2}}\xi-E}{A-C+D+B(2-4e^{2i\varphi})}, \\
			r_{a} &=\dfrac{2ie^{i\varphi}\sqrt{B}[\Delta\cos\varphi-\xi-\sqrt{(A+2B)}\sin\varphi]}{A-C+D+B(2-4e^{2i\varphi})}, \\
			t_{b}
			&=\dfrac{2i\sqrt{c_{2}}(\Delta-\xi\cos\varphi-2\sqrt{c_{1}}\sin\varphi e^{i\varphi})}{A-C+D+B(2-4e^{2i\varphi})}, \\
			r_{b}
			&=r_{a}.
		\end{align}
	\end{subequations}
	Here we introduce the variables
	\begin{eqnarray}
		A&=&c_{1}+c_{2},~~~B=\sqrt{c_{1}c_{2}},~~~C=\Delta^{2}-\xi^{2}, \nonumber \\
		D&=&2i(\sqrt{c_{1}}+\sqrt{c_{2}})(\Delta-e^{i\varphi}\xi), \nonumber \\
		E&=&c_{2}+(\sqrt{c_{1}}+i\Delta)^{2}+\xi^{2},
	\end{eqnarray}
	with $c_{1}=\lambda_{1}^{2}/\upsilon_{g}$ and $c_{2}=\lambda_{2}^{2}/\upsilon_{g}$. Here $\varphi_{j}\!=\!k_{j}L$ $(j=a,b)$ is the accumulated phase of the single photon when it passes through the coupling points of waveguide $j$. For simplicity, we assume $k_{i}\!=\!k$; then $\varphi_{j}\!=\!\varphi$. 
	According
	to the dispersion relation $\varepsilon\!=\!\upsilon_{g}k$,
	$\varphi$ can be written as a \ensuremath{\varepsilon}-dependent
	part plus a constant part, namely,
	\begin{eqnarray}
		\varphi\!&=&\!\tau\Delta+\theta,
	\end{eqnarray}
	with $\tau\!=\!L/\upsilon_{g}$ the propagating time
	of the photon between the two coupling points $x_{j_{1}}=0$ and $x_{j_{2}}=L$. We define $\gamma_{i}=\lambda_{i}/\upsilon_{g}~(i=1,2)$, which is the decay rate of the $i$th TLA. The important condition for distinguishing the Markovian regime from the non-Markovian one is whether the propagating time $\tau$ can be neglected. When the propagating time $\tau$  is much less than the relaxation time of the TLA, namely, $\tau\ll1/\gamma_{i}$ $\left(i=1,2\right)$, the propagating time $\tau$ can be neglected.
	Then $\varphi\simeq\theta$ and the system works in the Markovian regime. When the propagating time  $\tau$ is comparable to the relaxation time of the TLA, i.e., $\tau\sim1/\gamma_{i}$ $\left(i=1,2\right)$. The propagation time $\tau$ cannot be ignored and then the system works in the non-Markovian regime. Based on these amplitudes, both the reflection coefficient $R_{j}=|r_{j}|^{2}$ and the transmission coefficient $T_{j}=|t_{j}|^{2}$ can be obtained accordingly. We can check the relation $R_{a}+T_{a}+R_{b}+T_{b}=1$, which confirms the probability conservation.
	
	To study the nonreciprocal routing, we need to investigate the routing properties when the photon is input from another side of the system. As an example, we consider the photon input from port 2. Similarly, we can obtain the transmission and reflection amplitudes for the photon incident from port 2 as
	\begin{subequations}
		\begin{align}
			\widetilde{t}_{a}&=-\dfrac{2e^{2i\varphi}B+2ie^{i\varphi}\sqrt{c_{2}}\xi-2\sin\varphi\sqrt{c_{1}}\xi-E^{'}}{A-C+D+B(2-4e^{2i\varphi})}, \\
			\widetilde{r}_{a} &=\dfrac{2ie^{-i\varphi}\sqrt{B}[\Delta\cos\varphi-\xi-\sqrt{(A+2B)}\sin\varphi]}{A-C+D+B(2-4e^{2i\varphi})}, \\
			\widetilde{t}_{b}
			&=\dfrac{2i\sqrt{c_{1}}(\Delta-\xi\cos\varphi-2\sqrt{c_{2}}\sin\varphi e^{i\varphi})}{A-C+D+B(2-4e^{2i\varphi})}, \\
			\widetilde{r}_{b}
			&=\widetilde{r}_{a},
		\end{align}
	\end{subequations}
	where $ E^{'}=c_{1}+(\sqrt{c_{2}}+i\Delta)^{2}+\xi^{2}.$ Then both the reflection coefficient $\widetilde{R}_{j}=|\widetilde{r}_{j}|^{2}$ and the transmission coefficient $\widetilde{T}_{j}=|\widetilde{t}_{j}|^{2}$ can be obtained. We point out that both the transmission and reflection coefficients are valid in both the Markovian and non-Markovian regimes.
	Meanwhile, we introduce the chiral parameter 
	\begin{eqnarray}
		G=\gamma_{2}/\gamma_{1}.
	\end{eqnarray}
	The $G\!\neq\!1$ implies chirality in this system.
    When $G\!\!=\!\!1$, we find $T_{j}\!\!=\!\widetilde{T}_{j}~(j=a,b)$ and $R_{j}\!\!=\!\widetilde{R}_{j}$, and thus the scattering characteristics for the photon incident from either port 1 or port 2 is reciprocal. When $G\!\!\neq\!\!1$, we have $T_{j}\!\!\neq\!\widetilde{T}_{j}$ and $R_{j}\!\neq\!\widetilde{R}_{j}$.
	The routing properties for the photon incident from port 1 are different from those for the case of input from port 2. When the photon is input from one side of the waveguides, the coefficients for the photon output from each ports are different from the case for the photon injected from the other side of the waveguides. This indicates a nonreciprocity in this system. This result means that the chirality can induce a nonreciprocal routing.
	In the following, we discuss the single-photon nonreciprocal routing in both the Markovian and non-Markovian regimes.
	\vspace{-8pt}
	\section{SINGLE-PHOTON NONRECIPROCAL ROUTING IN THE MARKOVIAN REGIME}
	~\label{Markovian regime} 
	In this section, we discuss single-photon nonreciprocal routing in the Markovian regime, where the propagating time $\tau$ can be ignored, thus $\varphi\approx\theta$. This approximation is valid when the distance between the two TLAs is very short, i.e., $\tau\ll1/\gamma_{i}$ $\left(i=1,2\right)$. To study the nonreciprocal routing, we will investigate the single-photon routing process by considering the photon input from either port 1 or port 2. To obtain the scattering state of the photons, we calculate the transmission and reflection coefficients for each waveguide.
	\vspace{-10pt}
	\subsection{Single-photon nonreciprocal routing}
\begin{figure}[tbp]
		\center
		\includegraphics[width=8.5cm]{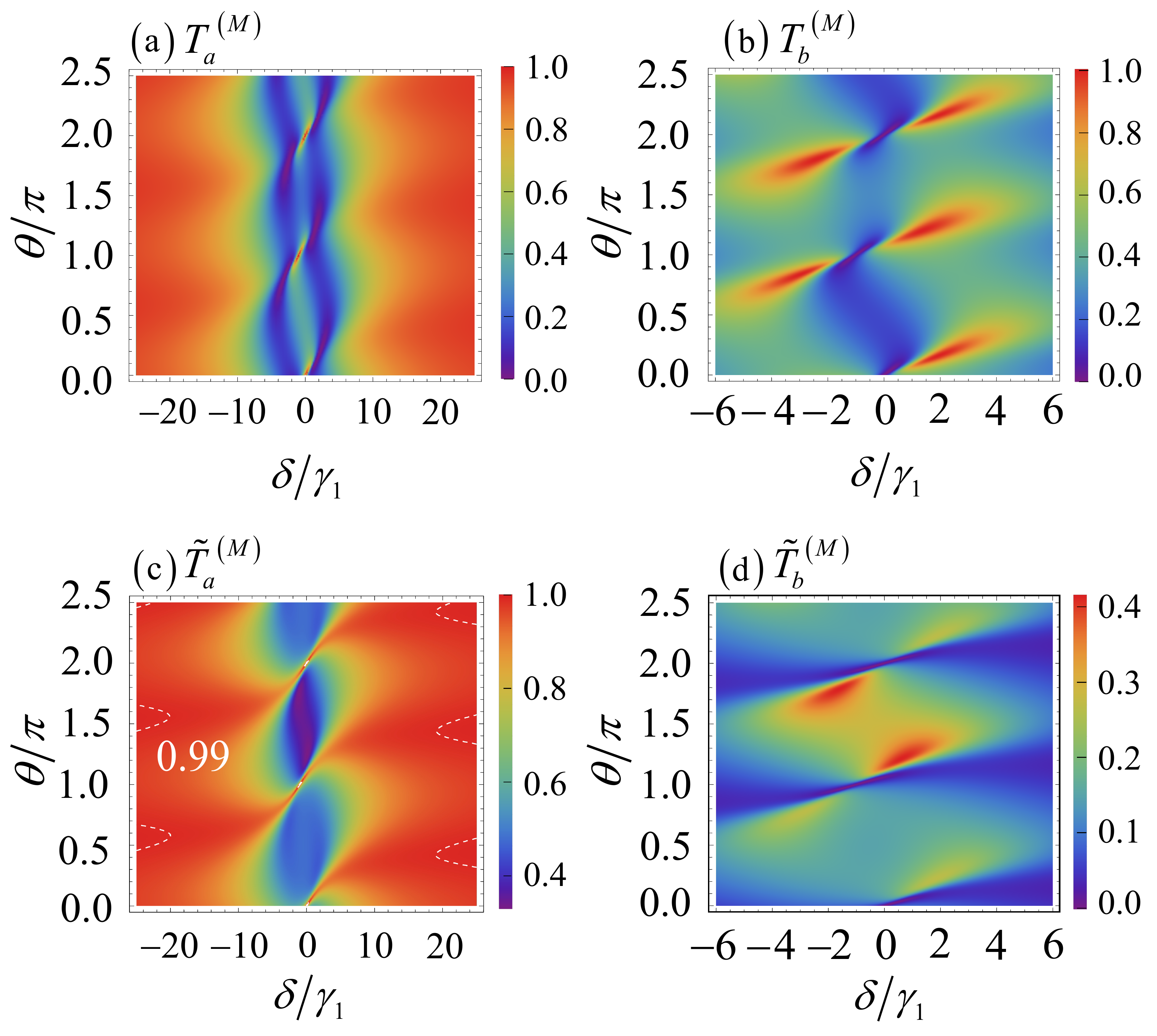}
		\caption{ Transmission coefficients (a) $T_{a}^{(M)}$ and (b) $T_{b}^{(M)}$ for the photon input from port 1 and (c) $\widetilde{T}_{a}^{(M)}$ and (d) $\widetilde{T}_{b}^{(M)}$ for the photon input from port 2 vs the detuning $\delta/\gamma_{1}$ and the  phase shift $\theta/\pi$ in the Markovian regime. The white dashed curve indicates $\widetilde{T}_{a}^{(M)}=0.99$. Other parameters are $G=2.38$ and $\xi/\gamma_{1}\!=\!0.5$.}
		\label{Fig2}
	\end{figure}

	We first explore the routing properties of a single photon input from port 1.~We plot the transmission coefficients $T_{j}^{\left(M\right)}$ $(j\!=\!a,b)$ as a function of the detuning  $\delta\!=\!\varepsilon-\left(E_{1}-E_{3}\right)$ and phase shift $\theta/\pi$ in Figs.~\ref{Fig2}(a) and~\ref{Fig2}(b), where $G=2.38$. 
	Here, to label the Markovian results, we add a superscript $(M)$ to $T_{j}$ . As shown in Figs.~\ref{Fig2}(a) and~\ref{Fig2}(b), we can observe that along the $\theta$-axis, $T_{j}^{\left(M\right)}$ is periodic with a period of $2\pi$. Besides, $T_{j}^{\left(M\right)}=1$ can be reached by choosing appropriate detuning $\delta/\gamma_{1}$ and phase shift $\theta$. This result means that the photon input from port 1 can be fully output from port 2 or 4. As shown in Fig.~\ref{Fig2}(a),~$T_{a}^{(M)}=1$ occurs at $(\delta/\gamma_{1}, \theta/\pi)\!=\!(-1,1),~(0,0)$, and $(0,2)$. 
	The phenomenon means that the photon input from port 1 is fully output from port 2. The $T_{a}^{(M)}\!=\!1$ at $\delta=0$ is just the resonance phenomenon that the incident photon is resonant with the transitions between the energy pairs $\left|E_{1}\right\rangle$ and $\left|E_{3}\right\rangle$, as well as $\left|E_{2}\right\rangle$ and $\left|E_{4}\right\rangle$, satisfying $\varepsilon\!=\!E_{1}-E_{3}=E_{2}-E_{4}$. Away from the resonance, namely, $\delta\neq0$, $T_{a}^{(M)}$ increases as $|\delta|$ increases. This is because when $\delta/\gamma_{1}$ increases to a sufficiently large value, the system enters a large detuned state and the single photon can be directly routed from port 1 to port 2. Besides the photon output from port 2, the photon can also be fully output from port 4. As shown in Fig.~\ref{Fig2}(b), it is found that $T_{b}^{(M)}\!\!=\!1$ at $(\delta/\gamma_{1}, \theta/\pi)=(-3.14,0.82)$ and $(1.5,1.22)$. This perfect routing $T_{b}^{(M)}\!\!=\!1$ is induced by the dipole coupling and the chirality. We point out that, the photon cannot be fully output from port 1 or 3. However, there is still some significant probability for the photon routing from port 1 to port 1 or 3 due to $R_{a}^{(M)}=R_{b}^{(M)}$.
	
	To understand the nonreciprocal properties, we investigate the routing process for the photon input from port 2. 
	We plot the transmission coefficient $\widetilde{T}_{j}^{(M)}$ as a function of $\ensuremath{\delta/\gamma_{1}}$ and $\theta/\pi$ in Figs.~\ref{Fig2}(c) and~\ref{Fig2}(d), where $G\!=\!2.38$.  It is found that the results in Figs.~\ref{Fig2}(c) and~\ref{Fig2}(d) are apparently different from those in Figs.~\ref{Fig2}(a) and~\ref{Fig2}(b), though they are obtained under the same system parameters. This is just the nonreciprocal routing. This nonreciprocal phenomenon is induced by the chirality. The photon can be perfectly routed from port 2 to port 1 at $(\delta/\gamma_{1},\theta/\pi)=(-1,1),~(0,0)$, and $(0,2)$. The $\widetilde{T}_{a}^{(M)}>0.99$ occurs in the range $19<\delta/\gamma_{1}<25$, $0.32<\theta/\pi<0.55$, and $1.34<\theta/\pi<1.57$ or $-25<\delta/\gamma_{1}<-20$, $0.46<\theta/\pi<0.67$, and $1.44<\theta/\pi<1.65$, as indicated by the white dashed curve in Fig.~\ref{Fig2}(c). We also note that $\widetilde{T}_{b}^{(M)}$ cannot reach 1 under the used parameters, which means that the photon input from port 2 cannot be fully output from port 3. This is different from Fig.~\ref{Fig2}(b), where ${T}_{b}^{(M)}$ can reach 1. If we choose $1/G=2.38$, $\widetilde{T}_{b}^{(M)}$ can reach 1, which means that the photon input from port 2 can be fully output from port 3. Meanwhile, some similar features also exist. For example, along the $\theta$ axis $\widetilde{T}_{j}^{(M)}$ is also periodic with a period of $2\pi$.
	\begin{figure}[t]
		\center
		\includegraphics[width=8.5cm]{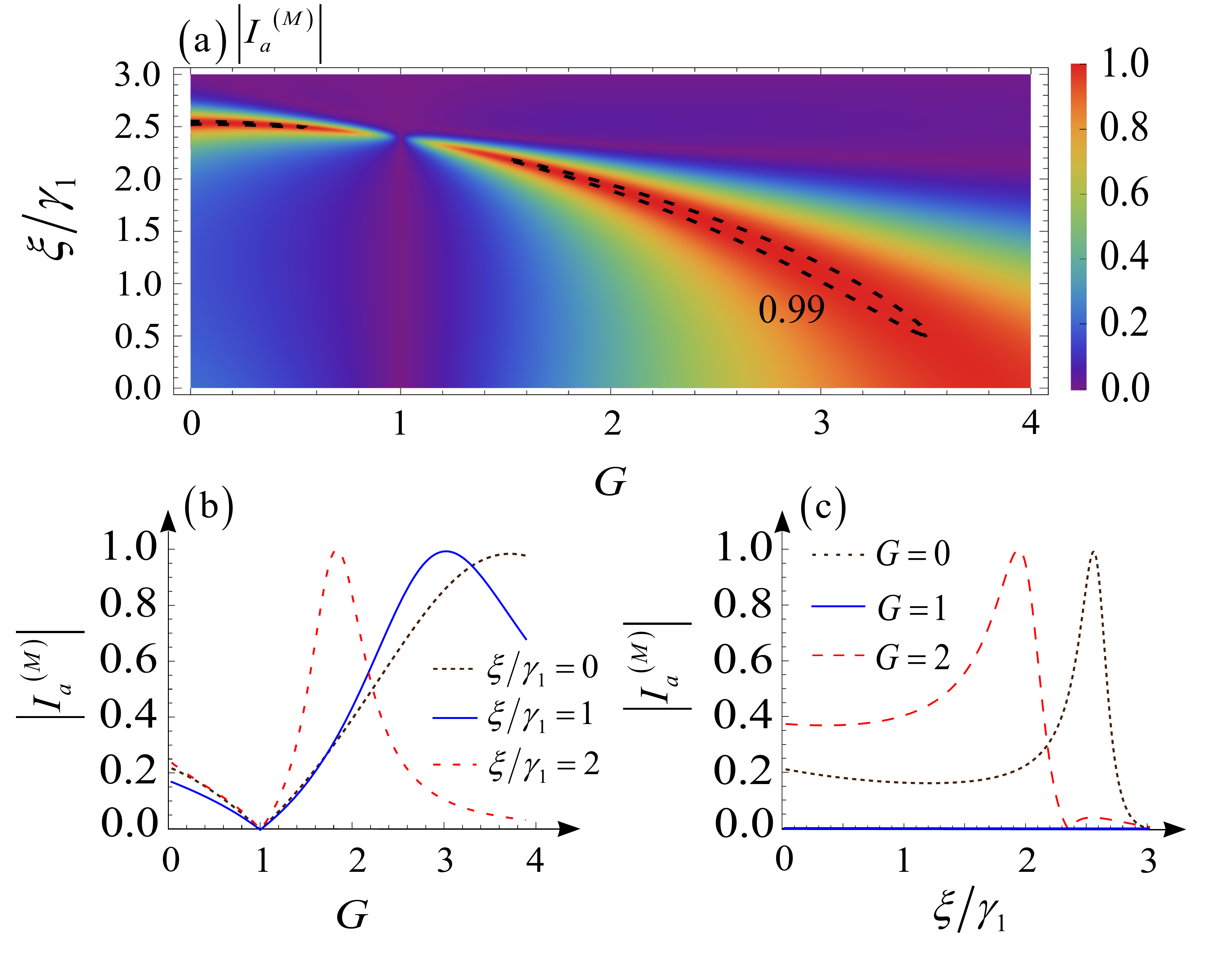}
		\caption{ (a) Contrast ratio  $\left|I_{a}^{\left(M\right)}\right|$ vs $G$ and $\xi/\gamma_{1}$. The  black solid curve indicates $\left|I_{a}^{\left(M\right)}\right|=0.99$. (b) Profiles of (a) at $\xi/\gamma_{1}=0$ (black dotted curve), $\xi/\gamma_{1}=1$ (blue solid curve), and $\xi/\gamma_{1}=2$ (red dashed curve). (c) Profiles of (a) at $G=0$ (black dotted curve), $G=1$ (blue solid curve), and $G=2$ (red dashed curve). The other common parameters used are $\theta/\pi=0.1,~\omega_{e}/\gamma_{1}=100$, and $\ensuremath{\varepsilon/\gamma_{1}}=103.$}
		\label{Fig3}
	\end{figure}
	
	To further analyze the reason for the appearance of single-photon nonreciprocal routing, we introduce the contrast ratio $I_{j}^{\left(M\right)}$, which is defined as
	\begin{eqnarray}
		I_{j}^{\left(M\right)}=\frac{\widetilde{T}_{j}^{\left(M\right)}-T_{j}^{\left(M\right)}}{\widetilde{T}_{j}^{\left(M\right)}+T_{j}^{\left(M\right)}},\hspace{0.5cm}j=a,b.
		\label{Eq16}
	\end{eqnarray}
	The $I_{j}^{(M)}\!\!\neq\!0$ implies a nonreciprocal transport.
	We mention that both $I_{a}^{(M)}$ and $I_{b}^{(M)}$ can characterize the nonreciprocal routing. We mainly focus on $I_{a}^{(M)}$ instead of $I_{b}^{(M)}$ to maintain conciseness, since the underlying physics is similar. To better illustrate which factors are involved in the single-photon nonreciprocal routing, we plot $\left|I_{a}^{\left(M\right)}\right|$ versus $G$
	and $\xi/\gamma_{1}$ in Fig.~\ref{Fig3}. As shown in Fig.~\ref{Fig3}(a), we find $\left|I_{a}^{(M)}\right|\!\!=\!0$ when $G\!\!=\!\!1$, which means that there is no nonreciprocal routing without chirality. With the help of chirality $\left(G\neq1\right)$, we find a broad range of nonreciprocal routing. As indicated by the black dashed curve, $\left|I_{a}^{(M)}\right|>0.99$ when $2.5<\xi/\gamma_{1}<2.58$, and $0<G<0.55$, or $0.5<\xi/\gamma_{1}<2.2$, and $1.47<G<3.52$.
	These results can also be observed from Figs.~\ref{Fig3}(b) and~\ref{Fig3}(c).
    Specifically, $\left|I_{a}^{(M)}\right|=1$ occurs at $(G,\xi/\gamma_{1})=(0,2.53),~(1.86,2),~(2,1.92)$,~and~$(3.1,1)$. The nonreciprocal routing is induced by the chirality. However, the suitable dipole coupling can enhance but not induce the nonreciprocal routing in the Markovian regime.
	\vspace{-7pt}
	\begin{figure}[tbp]
		\center
		\includegraphics[width=8.5cm]{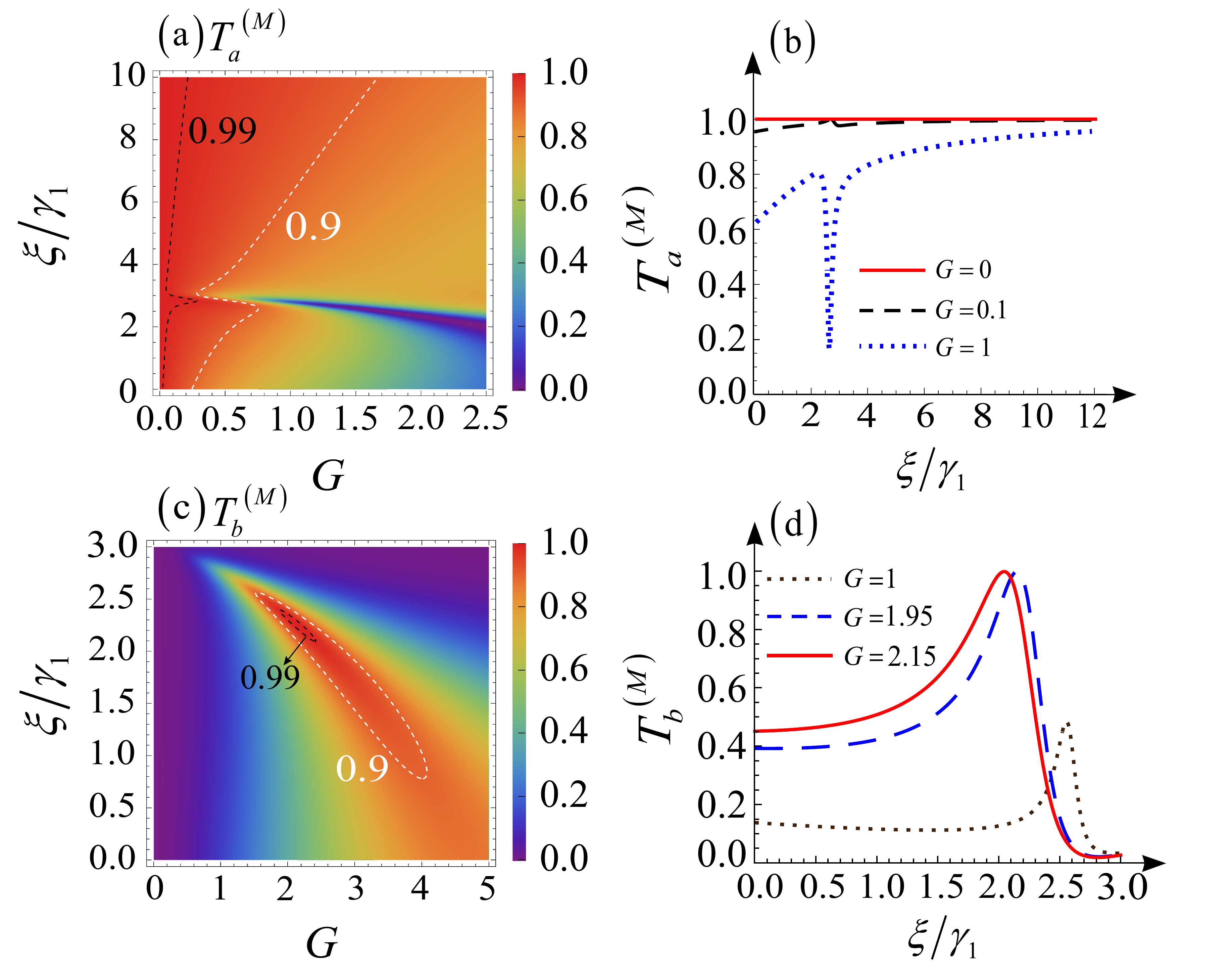}
		\caption{ Transmission coefficients (a) $T_{a}^{M}$ and (c) $T_{b}^{M}$ as a function of the chiral parameter $G$ and dipole coupling strength
			$\xi/\gamma_{1}$ for the photon  incident
			from port 1. The white and black dashed curves indicate $T_{j}^{\left(M\right)}=0.9$ and 0.99, respectively. The curves in (b) and (d) show the profiles of (a) and (c) at given values of $G$, respectively. The other common parameters used are $\theta/\pi=0.1$, $\ensuremath{\varepsilon/\gamma_{1}}=103.4$, and $\omega_{e}/\gamma_{1}=100.$}
		\label{Fig4}
	\end{figure}
	\vspace{-7pt}
	\subsection{Perfect single-photon routing}
	In this section we study the influence of the chirality and dipole coupling strength on the single-photon routing. To this end, we plot $T_{j}^{(M)}$ versus the chiral parameter $G$ and the dipole coupling strength $\xi/\gamma_{1}$ in Fig.~\ref{Fig4}. 
	According to Fig.~\ref{Fig4}(a), the single photon can be routed from port 1 to port 2 with $T_{a}^{(M)}>0.9$ when $0<G<1.67$ and $0<\xi/\gamma_{1}<10$, as indicated by the white dashed curve, while for $0<G<0.27$ and $0<\xi/\gamma_{1}<10$, the probability for the photon output from port 2 could be higher than 0.99, i.e., $T_{a}^{(M)}>0.99$. This parameter regime of almost perfect routing is indicated by the black dashed curve. To clearly see the dependence of $T_{a}^{(M)}$ on $G$ and $\xi/\gamma_{1}$, we plot $T_{a}^{(M)}$ versus $\xi/\gamma_{1}$ for $G\!=\!0,~0.1$, and 1 in Fig.~\ref{Fig4}(b).  $T_{a}^{(M)}\!=\!1$ occurs at $G\!=\!0$ regardless the value of $\xi$ and $(G,\xi/\gamma_{1})=(0.1,2.7)$. However, when $\xi/\gamma_{1}$ is large enough, it does not require $G\!=\!0$ to achieve $T_{a}^{(M)}\!=\!1$.  

	For the routing from port 1 to port 4, the perfect routing requires a larger $G$ compared with the results shown in Fig.~\ref{Fig4}(a). Figure~\ref{Fig4}(c) shows that $T_{b}^{(M)}>0.9$ occurs in the regime $0.78<\xi/\gamma_{1}<2.57$ and $1.52<G<4.02$, as indicated by the white dashed curve. A higher routing probability $T_{b}^{\left(M\right)}>0.99$ can be reached when $2.09<\xi/\gamma_{1}<2.39$ and $1.87<G<2.41$, as indicated by the black dashed curve. It is possible to perfectly route the photon from port 1 to port 4. To see clearly the dependence of $T_{b}^{\left(M\right)}$ on $G$ and $\xi/\gamma_{1}$, we plot $T_{b}^{\left(M\right)}$ versus $\xi/\gamma_{1}$ for $G=1, 1.95$, and 2.15 in Fig.~\ref{Fig4}(d).
	%, and vs $\xi/\gamma_{1}$ for $G=0, 1, 2$ in Fig.~\ref{Fig4} (f). 
	It can be seen that the single photon input from port 1 can perfectly output from port 4 at $(G,\xi/\gamma_{1})=(1.95,2.35)$~and~$(2.15,2.25)$, where $T_{b}^{\left(M\right)}=1$. Both the chirality and dipole coupling assist the perfect routing from port 1 to port 4. Without the help of chirality, $T_{b}^{\left(M\right)}$ cannot reach 1, as shown by the black dotted curve in Fig.~\ref{Fig4}(d).
	The above discussions demonstrates that the single-photon perfect routing can be realized by adjusting the chiral parameter $G$ and dipole coupling strength  $\xi/\gamma_{1}$. Our system provides a promising candidate for realizing a perfect single-photon router.
	\vspace{-15pt}
	\section{Single-photon nonreciprocal routing in the non-Markovian regime}~\label{non-Markovian regime}
	In the previous discussions, we have analyzed the single-photon nonreciprocal routing in the Markovian regime by neglecting the propagating time $\tau$. However, when the distance between the two TLAs is large enough, the propagating time $\tau$ is comparable to the lifetime $1/\gamma_{i}$ of the excited state $\left|e\right\rangle_{i} $ for the $i$th TLA, namely, $\tau\!\!\sim\!\!1/\gamma_{i}$ $\left(i\!=\!1,2\right)$. In this case, the system works in the non-Markovian regime. In this section, we study the single-photon nonreciprocal routing in the non-Markovian regime.
	\subsection{Single-photon nonreciprocal routing} 
\begin{figure}[t]
		\center
		\includegraphics[width=8.5cm]{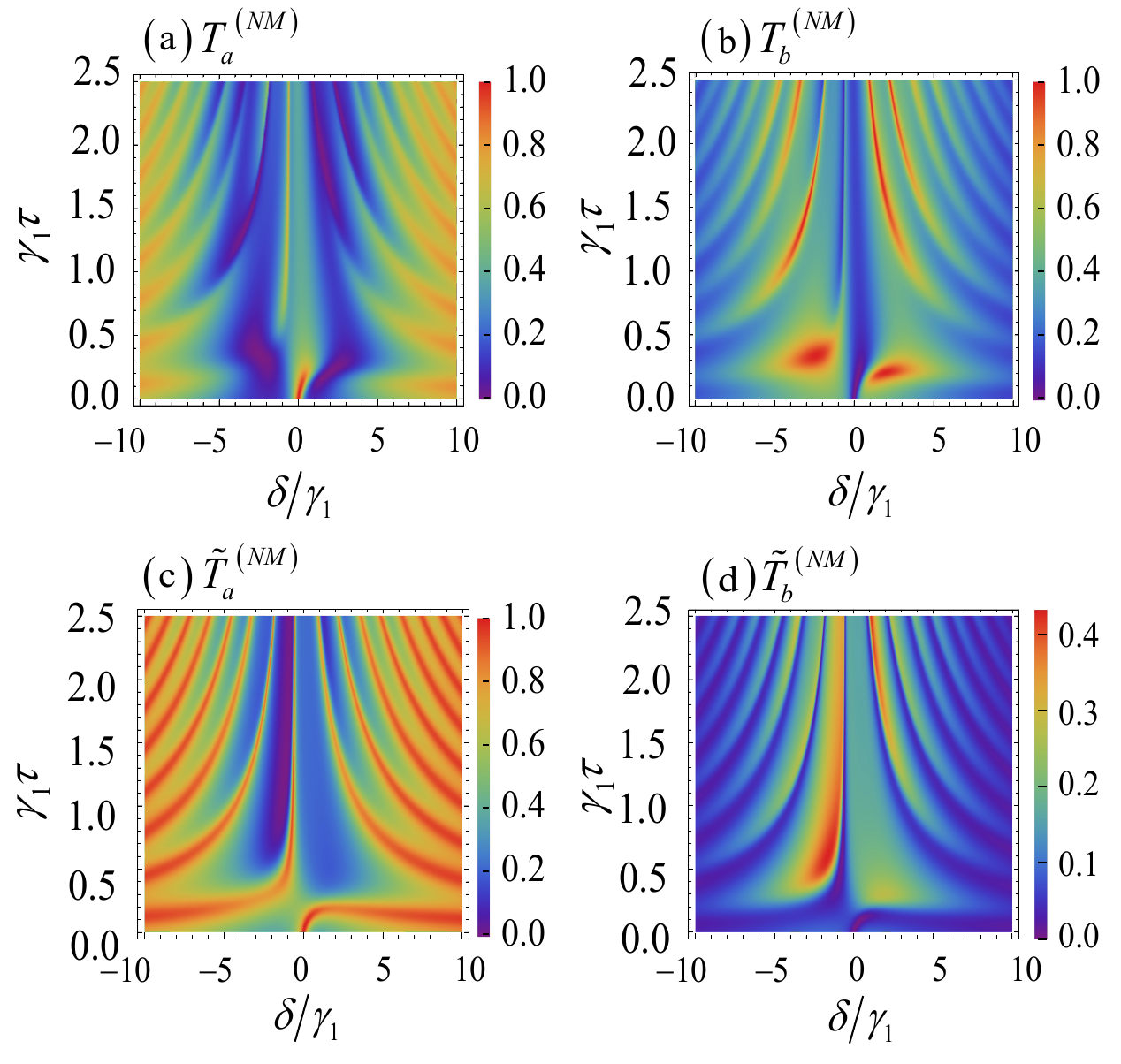}
		\caption{ Transmission coefficients (a) $T_{a}^{(N\!M)}$ and (b) $T_{b}^{(N\!M)}$ for the photon input from port 1 and (c) $\widetilde{T}_{a}^{(N\!M)}$ and (d) $\widetilde{T}_{b}^{(N\!M)}$  for the photon input from port 2 vs the detuning $\delta/\gamma_{1}$ and the scaled propagating time $\gamma_{1}\tau$. In all panels, the parameters are $G\!=\!2.38,~\xi/\gamma_{1}\!=\!0.5$, and  $\theta/\pi\!=\!0$.}
		\label{Fig5}
	\end{figure}

	We first explore the routing properties of the photon input from port 1 in the non-Markovian regime.  
	Here, to label the non-Markovian results, we add a superscript ($N\!M$) in all results, such as $T_{j}^{(N\!M)}$, $\widetilde{T}_{j}^{(N\!M)}$ and $I_{j}^{(N\!M)}$ $(j\!=\!a,b)$. In Figs.~\ref{Fig5}(a) and~\ref{Fig5}(b) we show the dependence of $T_{j}^{(N\!M)}$ on the detuning $\delta/\gamma_{1}$ and the scaled propagating time $\gamma_{1}\tau$. We find $T_{j}^{(N\!M)}$=1 under certain values of $\delta/\gamma_{1}$ and $\gamma_{1}\tau$, which means that the single photon input from port 1 can be fully output from ports 2 or 4 in the non-Markovian regime. Moreover, both $T_{a}^{(N\!M)}$ and $T_{b}^{(N\!M)}$ exhibit multiple peaks along the $\gamma_{1}\tau$ and $\delta/\gamma_{1}$ axes. This is induced by the non-Markovian effect.
	%%%%%%%%%%%%%%%%%%%%%%%%%%%%%%
	%%%%%%%%%%%%%%%%%%%%%%%%%%%%%% 
	As shown in Fig.~\ref{Fig5}(a), we find that $T_{a}^{(N\!M)}\!\!=\!1$ for $\delta=\!0$ and $\gamma_{1}\tau=\!0$. When $\gamma_{1}\tau\neq0$, $T_{a}^{(N\!M)}$ is reduced at $\delta=0$. For $|\delta/\gamma_{1}|<10$, $\gamma_{1}\tau\neq0$, $T_{a}^{(N\!M)}<1$. However, $T_{a}^{(N\!M)}$ increases as $|\delta/\gamma_{1}|$ increases. When $|\delta/\gamma_{1}|>47$, we have $T_{a}^{(N\!M)}\!>\!0.99$. For $T_{b}^{(N\!M)}$, we find $T_{b}^{(N\!M)}=1$ at $(\delta/\gamma_{1}, \gamma_{1}\tau)=(1.86,0.2),~(-2.33,0.35)$, and $(-2.92,1.5)$. The $T_{b}^{(N\!M)}$ decreases with some oscillations as $|\delta/\gamma_{1}|$ increases. We also note that along the $\gamma_{1}\tau$ axis, $T_{j}^{(N\!M)}$ $(j\!=\!a,b)$ periodically oscillates with a period of $2\pi\gamma_{1}/\Delta$, showing multiple peaks.
	Specifically, $T_{b}^{(N\!M)}$ can reach 1 at $(\delta/\gamma_{1}, \gamma_{1}\tau)=(-3,1.47),~(1.5,1.91)$, and $(2,0.21)$. Thus, the distance between the two TLAs can assist the photon routing from port 1 to port 4.
	The above discussions demonstrate that it is possible to realize perfectly a directional single-photon routing in the non-Markovian regime. We point out that there is no determined boundary between the Markovian and non-Markovian regimes. The behavior of the single-photon routing varies continuously when the system transits from the Markovian regime to the non-Markovian regime.
	
	To study the nonreciprocal routing in the non-Markovian regime, we consider the photon input from port 2 and show the results of transmission coefficients $\widetilde{T}_{j}^{(N\!M)}$ $(j\!=\!a,b)$ in Figs.~\ref{Fig5}(c) and~\ref{Fig5}(d) with the same parameters as that in Figs.~\ref{Fig5}(a) and~\ref{Fig5}(b). By comparing with Figs.~\ref{Fig5}(a) and~\ref{Fig5}(b), we find that the characteristics of $\widetilde{T}_{j}^{(N\!M)}$ are obviously different from ${T}_{j}^{(N\!M)}$, though the system parameters are the same. This is a nonreciprocal routing phenomenon. We also find $\widetilde{T}_{a}^{(N\!M)}>T_{a}^{(N\!M)}$ but $\widetilde{T}_{b}^{(N\!M)}<T_{b}^{(N\!M)}$ in most parameter regimes. The photon input from port 2 is more easily to be output from waveguide $a$ than input from port 1. However, when the photon is expected to  output from waveguide $b$, it is better to inject the photon from port 1 of waveguide $a$. Despite the difference, there are still some similar features. For instance, $\widetilde{T}_{j}^{(N\!M)}$ still oscillates with $\gamma_{1}\tau$ and $\delta/\gamma_{1}$ at large $\gamma_{1}\tau$. Along the $\gamma_{1}\tau$ axis, $\widetilde{T}_{j}^{(N\!M)}$ also periodically oscillates with a period of $2\pi\gamma_{1}/\Delta$. We point out that, $\widetilde{T}_{a}^{(N\!M)}\!=\!1$ occurs at $(\delta/\gamma_{1},\gamma_{1}\tau)\!=\!(0,0)$, and when $|\delta/\gamma_{1}|>20$, $\widetilde{T}_{a}^{(N\!M)}>0.99$. However $\widetilde{T}_{b}^{(N\!M)}$ cannot reach 1 for $\gamma_{1}\tau<2.5$, $|\delta/\gamma_{1}|<10$, and $\gamma_{2}/\gamma_{1}=2.38$. If $\gamma_{1}/\gamma_{2}=2.38$, $\widetilde{T}_{b}^{(N\!M)}$ can reach 1, which means that the photon input from port 2 can be fully output from port 3 in the non-Markovian regime.
\begin{figure}[tbp]
	\center
	\includegraphics[width=8.5cm]{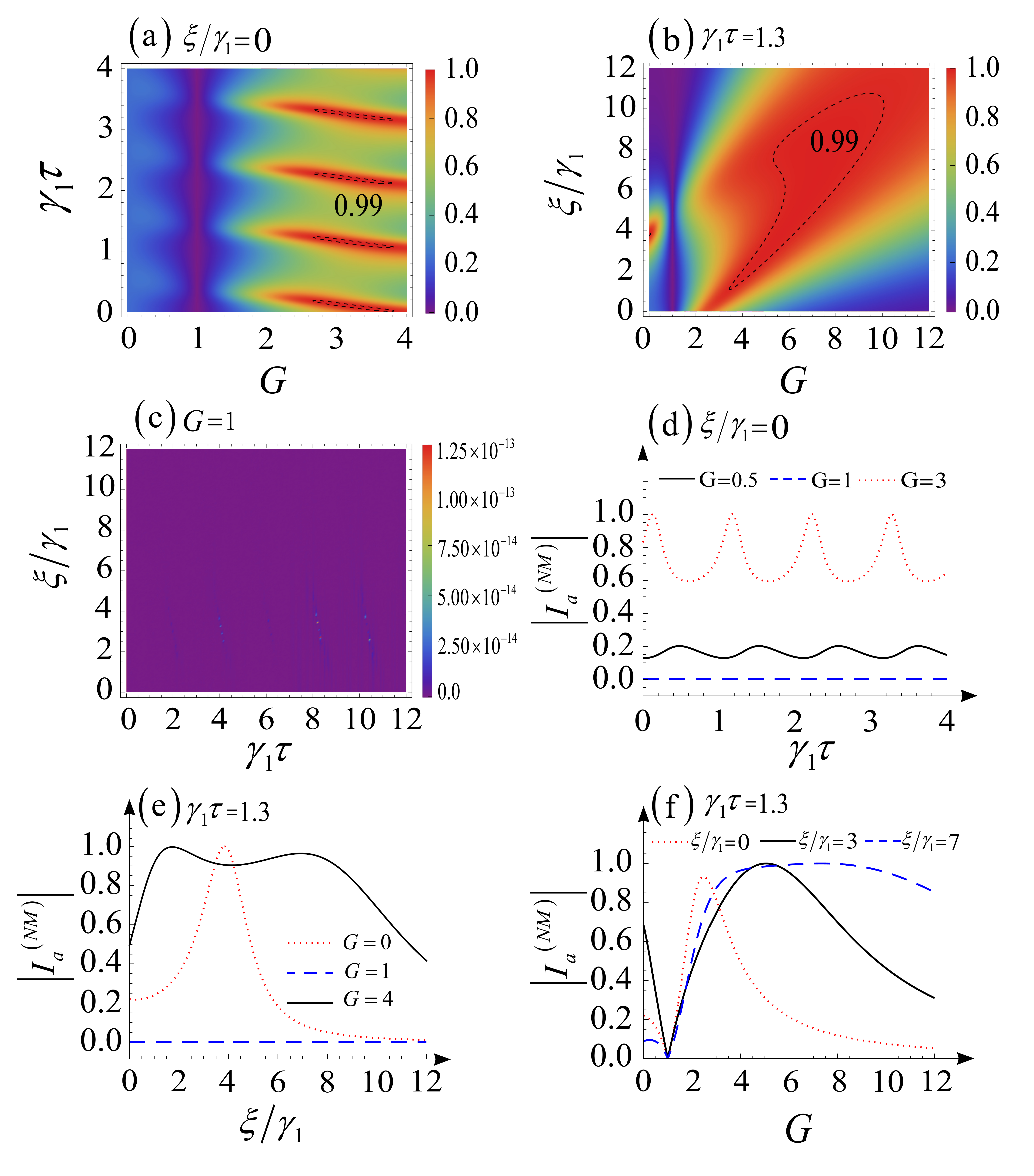}
	\caption{ (a) Contrast ratio $\left|I_{a}^{(N\!M)}\right|$ vs the scaled propagating time $\gamma_{1}\tau$ and chiral parameter $G$ for $\xi/\gamma_{1}=0$. (b) $\left|I_{a}^{(N\!M)}\right|$ vs $\xi/\gamma_{1}$ and $G$ for $\gamma_{1}\tau=1.3$. The black dashed curve in (a) and (b) indicates $\left|I_{a}^{(N\!M)}\right|=0.99$. (c) Contrast ratio $\left|I_{a}^{(N\!M)}\right|$ vs $\gamma_{1}\tau$ and $\xi/\gamma_{1}$ for $G=1$. The curves in (d) show the profiles of (a) at $G=0.5$ (black solid curve), $1$ (blue dashed curve), and $3$ (red dotted curve) for $\xi/\gamma_{1}=0$. (e) Profiles of (b) at given value of $G=0$ (red dotted curve), 1 (blue dashed curve), and 4 (black solid curve) for $\gamma_{1}\tau=1.3$. (f) Profiles of (b) at given value of $\xi/\gamma_{1}=0$ (red dotted curve), 3 (black solid curve), and 7 (blue dashed curve) for $\gamma_{1}\tau=1.3$. The other common parameters used are $\theta/\pi=0.1$, $\ensuremath{\varepsilon/\gamma_{1}}=103$, and $\omega_{e}/\gamma_{1}=100.$}
	\label{Fig6}
\end{figure}
To further understand the physical reason of the nonreciprocal routing in the non-Markovian regime, we study the contrast ratio $I_{j}^{(N\!M)}$, which can be obtained by replacing $\widetilde{T}_{j}^{(M)}$ and $T_{j}^{(M)}$ with $\widetilde{T}_{j}^{(N\!M)}$ and $T_{j}^{(N\!M)}$ in Eq.~(\ref{Eq16}), respectively. We show the results in Fig.~\ref{Fig6}. In Fig.~\ref{Fig6}(a) we plot $|I_{a}^{(N\!M)}|$ versus $G$ and $\gamma_{1}\tau$ for $\xi/\gamma_{1}=0$.   It is shown that $\left|I_{a}^{(N\!M)}\right|\!=\!0$ when $G\!=\!1$ regardless of $\gamma_{1}\tau$, which also can be observed by the dashed curve in Fig.~\ref{Fig6}(d). This result means that the non-Markovian effect cannot independently induce the single-photon nonreciprocal routing. We also find that $I_{a}^{(N\!M)}$ oscillates along the $\gamma_{1}\tau$ axis with a period of $\pi\gamma_{1}/\Delta$ for $G\neq0$ and 1, which can be seen clearly by the dotted curve in Fig.~\ref{Fig6}(d). The nonreciprocity in the regime $G>1$ could be higher than that in the regime $0<G<1$. This result can also be verified by Fig.~\ref{Fig6}(d). The black dashed curve in Fig.~\ref{Fig6}(a) indicates $\left|I_{a}^{(N\!M)}\right|=0.99$. It is found that $\left|I_{a}^{(N\!M)}\right|>0.99$ when $2.7<G<3.8$, $0.02<\gamma_{1}\tau<0.18$ and $1.05<\gamma_{1}\tau<1.25$ though $\xi/\gamma_{1}=0$. For example, $\left|I_{a}^{(N\!M)}\right|$ can reach 1 for $G\!=\!3$ at $\gamma_{1}\tau$=1.19 and 2.21, as shown by the dotted curve in Fig.~\ref{Fig6}(d). The chirality can assist the nonreciprocal routing in the non-Markovian regime. Figure~\ref{Fig6}(b) shows the contrast $|I_{a}^{(N\!M)}|$ ratio versus $G$ and $\xi/\gamma_{1}$ for $\gamma_{1}\tau=1.3$. We still find $|I_{a}^{(N\!M)}=0|$ at $G=1$ regardless of the value of $\xi/\gamma_{1}$ in the non-Markovian regime. This result indicates that there is no nonreciprocity in the non-Markovian regime even though the dipole coupling exists. With the help of chirality $\left(G\neq1\right)$, we can find a broad range of nonreciprocal routing. The black dashed curve indicates $\left|I_{a}^{(N\!M)}\right|=0.99$. Specifically, $\left|I_{a}^{(N\!M)}\right|>0.99$ when $3.4<G<10.1$ and $0.8<\xi/\gamma_{1}<10.7$. Thus the dipole coupling can enhance the nonreciprocal routing. 

To see clearly these features, we show the profiles of Fig.~\ref{Fig6}(b) at $G=0$, 1, and 4 in Fig.~\ref{Fig6}(e) and at $\xi/\gamma_{1}=0$, 3, and 7 in Fig.~\ref{Fig6}(f). 
We find that $\left|I_{a}^{(N\!M)}\right|=1$ at $(G,~\xi/\gamma_{1})=(0,~3.84)$ and $(4,1.7)$ in Fig.~\ref{Fig6}(e). As shown in Fig.~\ref{Fig6}(f), $\left|I_{a}^{(N\!M)}\right|$ reduces to 0 as $G$ increases from 0 to 1, followed by a peak as $G$ further increases from 1. In addition, as $\xi/\gamma_{1}$ increases, the location of the peak moves along the $G$ axis. This phenomenon indicates that a larger chirality requires a larger dipole coupling to reach the maximum nonreciprocal routing in the non-Markovian regime. The result in Fig.~\ref{Fig6}(b) is completely different from the Markovian result shown in Fig.~\ref{Fig3}. To see the influence of the dipole coupling and the non-Markovianity on the nonreciprocity, we plot $\left|I_{a}^{(N\!M)}\right|$ versus $\gamma_{1}\tau$ and $\xi/\gamma_{1}$ at $G=1$ in Fig.~\ref{Fig6}(c). The result $\left|I_{a}^{(N\!M)}\right|\!=\!0$ means that there is no nonreciprocal routing without chirality in the non-Markovian regime even though both the dipole coupling and the non-Markovianity exist. 
\begin{figure}[t]
	\center
	\includegraphics[width=8.5cm]{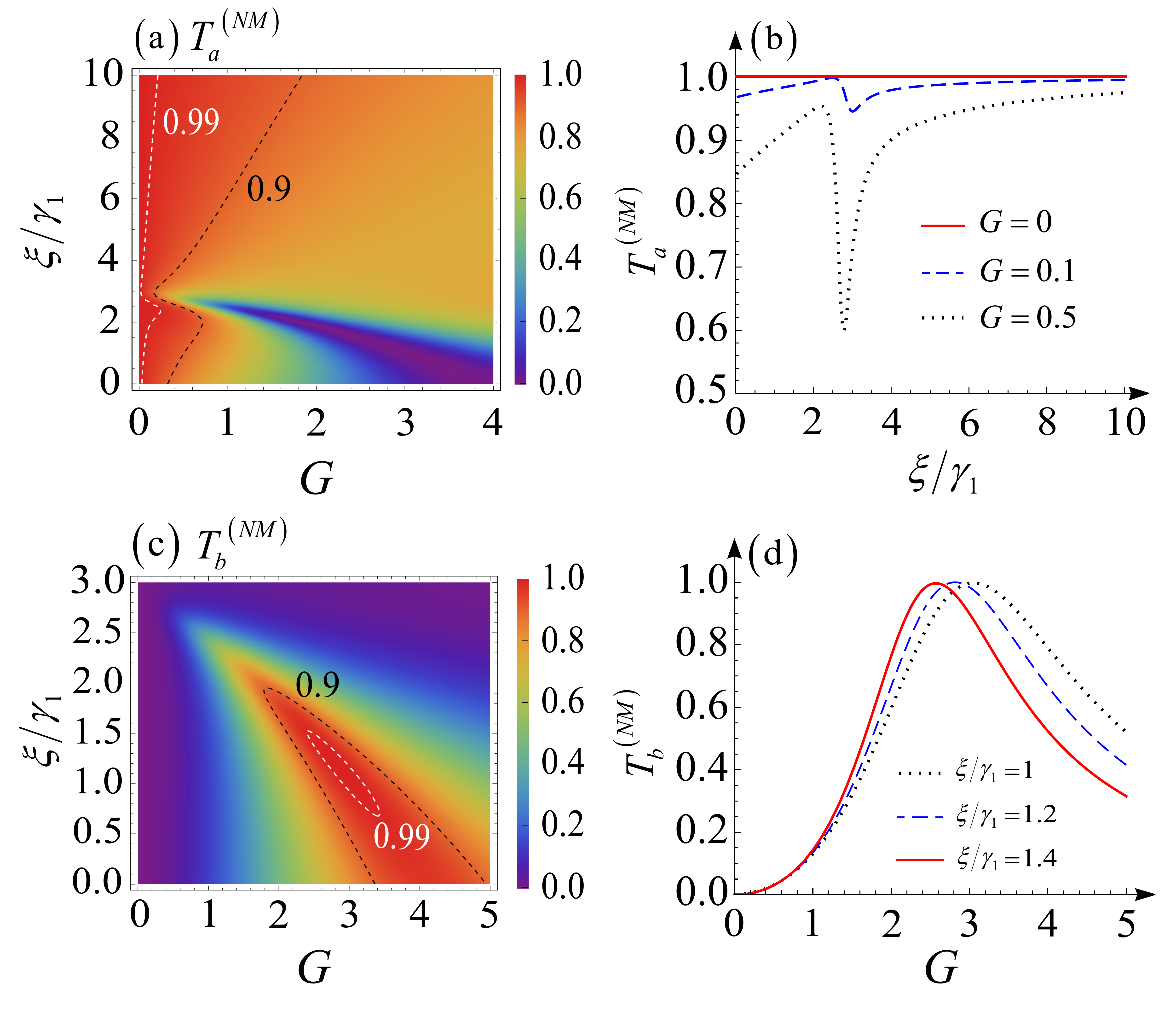}
	\caption{ Transmission coefficients (a) $T_{j}^{(N\!M)}$ and (c) $T_{b}^{(N\!M)}$
		vs the chiral parameter $G$ and dipole coupling strength
		$\xi/\gamma_{1}$ for the single photon  incident
		from port 1. The black and white dashed curves indicate $T_{j}^{(N\!M)}\!\!=\!0.9$ and 0.99, respectively. (b) Scattering coefficients $T_{a}^{(N\!M)}$ vs $\xi/\gamma_{1}$ for $G\!=\!0$ (red solid curve), $0.1$ (blue dashed curve), and $0.5$ (black dotted curve). (d) Scattering coefficient $T_{b}^{(N\!M)}$ vs $G$ for $\xi/\gamma_{1}\!=\!1$ (black dotted curve), $1.2$ (blue dashed curve), and $1.4$ (red solid curve). The other common parameters are $\theta/\pi\!=\!0.1$, $\ensuremath{\varepsilon/\gamma_{1}}\!=\!103.4$, $\gamma_{1}\tau\!=\!1.9$, and $\omega_{e}/\gamma_{1}\!=\!100.$}
	\label{Fig7}
\end{figure}
\vspace{-5pt}
\subsection{Perfect single-photon routing} 
In this section, we would like to discuss the influence of both the dipole coupling and the chirality on the single-photon routing in the non-Markovian regime.
To this end, we plot $T_{j}^{(N\!M)}$ $\left(j=a,b\right)$
versus the chiral parameter $G$ and the dipole coupling strength $\xi/\gamma_{1}$ in Fig.~\ref{Fig7}. According to Fig.~\ref{Fig7}(a), the single photon routed from port 1 to port 2 with probability $T_{a}^{(N\!M)}>0.9$ requires $0<G<1.82$, as indicated by the black dashed curve. Importantly, when $0<G<0.2$, $T_{a}^{(N\!M)}$ could be higher than 0.99, as indicated by the white dashed curve. Both the chirality and the dipole coupling can assist the photon routing from port 1 to port 2. To see this point more clearly, Fig.~\ref{Fig7}(b) shows the profiles of Fig.~\ref{Fig7}(a) at $G\!=\!0, 0.1$, and 0.5. We find $T_{a}^{(N\!M)}\!=\!1$ for $G\!=\!0$. When $G\!=\!0.1$, $T_{a}^{(N\!M)}\!=\!1$ occurs at $\xi/\gamma_{1}\!=\!2.47$. These indicate that both the chirality and the dipole coupling can assist the photon perfect routing from port 1 to port 2. We plot $T_{b}^{(N\!M)}$ versus $G$ and $\xi/\gamma_{1}$ in Fig.~\ref{Fig7}(c). As shown in Fig.~\ref{Fig7}(c), $T_{b}^{(N\!M)}>0.99$ when $0.68<\xi/\gamma_{1}<1.53$ and $2.4<G<3.4$, as indicated by the white dashed curve. To see clearly the dependence of $T_{b}^{(N\!M)}$ on $G$, we plot $T_{b}^{(N\!M)}$ versus $G$ for  $\xi/\gamma_{1}\!=\!1, 1.2$, and $1.4$ in Fig.~\ref{Fig7}(d). We observe $T_{b}^{(N\!M)}\simeq1$ at the points $(G,\xi/\gamma_{1})=(3.05,1),(2.81,1.2)$, and $(2.58,1.4)$.
It can also be seen that, as the dipole coupling strength $\xi/\gamma_{1}$ increases, the location of the peak moves towards $G=0$. The increasing dipole coupling can compensate for the reduction of the chirality to achieve a perfect photon routing from port 1 to port 4. Moreover, compared with the Markovian results in Fig.~\ref{Fig4}, the parameter range for $T_{b}>0.99$ is obviously enlarged due to the non-Markovianity. Thus, the non-Markovianity can assist the photon routing from port 1 to port 4. 
In the non-Markovian regime, we can still achieve a perfect single-photon routing by choosing appropriate dipole coupling strength $\xi$ and chiral parameter $G$. Moreover, the non-Markovianity can enhance the photon routing from waveguide $a$ to waveguide $b$. Therefore, the nonreciprocal routing is induced by the chirality, and both the dipole coupling and the non-Markovian effect can enhance the nonreciprocity.
\vspace{-13pt}
\section{DISCUSSIONS}~\label{DISCUSSIONS}
%Finally, we present some discussions on the experimental implementation of this scheme. In our scheme, there are two key elements for implementation of the waveguide-QED system. One is the realization of the chiral coupling between the TLA and the waveguide, and the other is the realization of the dipole interaction between the two TLAs. Therefore, the platform candidates suggested to implement the present scheme should be able to realize these two couplings. 
We now discuss the experimental implementation of this
scheme. In our scheme, there are two key elements for
implementation of the waveguide-QED system. One is the
realization of the chiral coupling between the TLA and the
waveguide, and the other is the realization of the dipole
interaction between the two TLAs. Therefore, the platform
candidates suggested to implement the present scheme should
be able to realize these two couplings.Currently, the chiral atom-field interactions have been realized in several experimental systems, including a gold nanoparticle~\cite{Petersen2014,coles2016} or cesium atoms~\cite{Mitsch2014} interacting with a tapered nanofiber, semiconducting quantum dots coupled to photonic crystal waveguides~\cite{Söllner2015}, and superconducting qubits coupled to transmission line waveguide~\cite{hoi2011demonstration,Barends2014,Astafiev2010,vanLoo2013,Liu2017}. Here we suggest to implement our scheme with the circuit waveguide-QED platform. Recently, it has been suggested to realize the chiral couplings by utilizing superconducting qubits coupled to a transmission line with circulators inserted to provide the chirality~\cite{Gu2017}. In addition, the dipole coupling between two superconducting qubits has been realized in experiments~\cite{DiCarlo2009, Chen2014,yan2018,Blais2021}. In some artificial systems, the dipole coupling can be introduced through other mediate systems or special design. For example, in superconducting circuits, the coupling of two two-level artificial atoms can be achieved through introducing a capacitance. Due to the adjustability of capacitive coupling, the coherent coupling strength can be adjusted~\cite{Zhong2021}. Moreover, the dissipative coupling can be achieved by capacitantly coupling two transmon qubits to a meandering transmission line~\cite{ren2025}. By adjusting the transmission line, the atomic spacing can be adjusted so that the system can work in the non-Markovian regime. These analyses indicate that the experimental implementation of the present scheme should be accessible with current and the near future experimental conditions.

We also mention the advantages of the present work compared with some previous single-atom waveguide-QED schemes. (i) The dual-waveguide dual-atom four-port system considered in this work not only supports the possibility of multipath photon routing, but also provides the alternative control means for realizing the optimization of photon routing efficiency through the controlled emitter frequencies and dipole-dipole coupling. (ii) Our scheme achieves nonreciprocal routing by combining the physical effects of chiral coupling and coupling configurations, which exhibits a new combined physical mechanism for single-photon routing. (iii) The exact analytical expressions of the single-photon scattering amplitude have been obtained using the real-space method, and the physical behaviors in both the Markovian and non-Markovian regimes have been analyzed. As a result, these detailed results will be helpful to the design of realistic physical systems.

\section{CONCLUSION}~\label{Conclusion}
We have proposed a scheme to implement a nonreciprocal single-photon router with a waveguide-QED platform. This system is composed of two dipole coupled two-level atoms, which are chirally coupled to two one-dimensional infinite waveguides. %\textcolor{blue}{The controllability of this system has been enhanced, supported by multiple tuning parameters, including the coupling strength between atoms and between atoms and waveguides, as well as the detuning between incident photons and energy levels. These parameters collectively make the routing control more flexible and robust.} 
We have studied the single-photon routing processes in both the Markovian and non-Markovian regimes. In the Markovian regime, we have found that the chiral coupling ($G\neq1$) between the two-level atoms and the waveguides is necessary to realize nonreciprocal routing in this system and that the dipole coupling between the two-level atoms can enhance the nonreciprocal routing. In order to achieve perfect single-photon directional routing, such as transmission from waveguide $a$ to waveguide $b$ or complete transmission in waveguide $a$, both the chiral coupling conditions and dipole coupling must be satisfied. In the non-Markovian regime, we have found that the realization of nonreciprocal routing also requires chiral coupling between the two-level atoms and the waveguides and that both the non-Markovian effect and the dipole coupling between the two-level atoms can enhance the nonreciprocal routing. To achieve perfect single-photon directional routing, such as transmission from waveguide $a$ to waveguide $b$ or complete unidirectional transmission in waveguide $a$, it is imperative to satisfy the conditions of both the chiral coupling and the presence of the dipole coupling. 
%\textcolor{blue}{Notably, it can maintain high routing performance even under non-ideal chiral coupling strength, significantly easing the limitations of ideal transmitter positioning or dedicated waveguide design. We have noticed that the actual implementation will involve additional challenges such as pulse distortion, noise and measurement error. A comprehensive analysis of these realistic conditions constitutes the basic direction of future research.}
\vspace{5pt}
\begin{acknowledgments}
	J.-Q.L. was supported in part by National Key Research and Development Program of China (Grant No.
	2024YFE0102400), National Natural Science Foundation of
	China (Grants No. 12175061, No. 12575015, No. 12247105,
	No. 11935006, and No. 12421005), and Hunan Provincial
	Major Sci-Tech Program (Grant No. 2023ZJ1010). J.-F.H.
	was supported in part by National Natural Science Foundation
	of China (Grants No. 12075083 and No. 12475016), Young
	Talents of the Furong Program (Grants No. 2025RC3141) and
	Key Program of Xiangjiang-Laboratory in Hunan province,
	China (Grant No. XJ2302001).
	%J.-F.H. is supported, in part, by the National Natural Science Foundation of China~(Grants No. 12075083 and No. 12475016) and the Key Program of Xiangjiang-Laboratory in Hunan province, China (Grant No. XJ2302001).
\end{acknowledgments}

\end{document}